\documentclass[11pt]{article}
\usepackage{epsfig}
\input{epsf}
\begin{document}

\newcommand{\beq}{\begin{equation}}
\newcommand{\eeq}{\end{equation}}
\newcommand{\beqa}{\begin{eqnarray}}
\newcommand{\eeqa}{\end{eqnarray}}
\newcommand{\ve}{\varepsilon}
\newcommand{\krig}[1]{\stackrel{\circ}{#1}}
\newcommand{\barr}[1]{\not\mathrel #1}
\newcommand{\vs}{\vspace{-0.003cm}}
\newcommand{\dfrac}{\displaystyle \frac}

\begin{flushright}
{\tiny{HISKP-TH-06/36}}
\end{flushright}

\vspace{.6in}

\begin{center}

{{\huge\bf Chiral perturbation theory}}

\end{center}

\vspace{.0in}

\begin{center}

{\Large
V\'eronique Bernard$^\star$\footnote{email: bernard@lpt6.u-strasbg.fr},
Ulf-G. Mei{\ss}ner$^\ddagger$$^\ast$\footnote{email: meissner@itkp.uni-bonn.de}
}

\vspace{1cm}

$^\star${\it Universit\'e Louis Pasteur, Laboratoire de Physique
            Th\'eorique\\ 3-5, rue de l'Universit\'e,
            F--67084 Strasbourg, France}

\bigskip

$^\ddagger${\it Universit\"at Bonn,
Helmholtz--Institut f\"ur Strahlen-- und Kernphysik (Theorie)\\
Nu{\ss}allee 14-16,
D-53115 Bonn, Germany}

\bigskip

 $^\ast${\it Forschungszentrum J\"ulich, Institut f\"ur Kernphysik
(Theorie)\\ D-52425 J\"ulich, Germany}

\bigskip

\bigskip

\end{center}

\vspace{.2in}

\thispagestyle{empty}

\begin{abstract}
\noindent
We give a brief introduction to chiral perturbation theory in its various
settings. We discuss some applications of recent interest including chiral 
extrapolations for lattice gauge theory.
\end{abstract}

\bigskip
\bigskip

\begin{center}
\end{center}

\bigskip
\bigskip

\centerline{Commissioned article for {\em Ann. Rev. Nucl. Part. Sci.}}
\vfill\eject


\tableofcontents

\section{Introduction and disclaimer}
\label{sec:intro}

Quantum chromodynamics (QCD) is the theory of the strong interactions.
While asymptotic freedom allows for a perturbative analysis at large
energies,  the low energy domain is characterized by the appearance of
hadrons that contain (and hide) the fundamental QCD degrees of freedom 
-- the quarks and gluons. This property of QCD is called {\em confinement}, 
it is beyond the reach of perturbation theory, calling  for a 
non-perturbative treatment.
Furthermore, QCD also exhibits spontaneous, explicit and anomalous
symmetry breaking -- and exactly the consequences of these broken
symmetries can be analyzed in terms of an appropriately formulated
effective field theory (EFT). This EFT is chiral perturbation theory
(CHPT) and in the following we will review its salient properties 
together with some phenomenological applications and the connection
to the lattice formulation of QCD.  Lattice QCD promises exact 
solutions utilizing a formulation on a discretized space-time and 
solving the pertinent path integral with the help of large computers.
Most lattice calculations are, however, done at unphysically large quark
masses -- and chiral perturbation theory offers a model-independent scheme
to perform the necessary chiral extrapolations.  

We end this introduction with a {\bf disclaimer}: This is not an all purpose 
review but rather stresses some fundamentals and selected applications. In
what follows, we supply a sufficient amount of references for the reader to
immerse deeper into the subject.

The manuscript is organized as follows. In Sec.~\ref{sec:symm} we discuss the
symmetries of QCD and their realization underlying the effective field theory.
The corresponding effective Lagrangian and the pertinent power counting
are given in Sec.~\ref{sec:Leff}, while Sec.~\ref{sec:loops} contains 
some remarks on chiral loops and the meaning of the low--energy coupling
constants of the EFT. As a specific example, the scalar form factor of
the pion is analyzed to one loop accuracy in Sec.~\ref{sec:sff}. The role 
of unitarity and analyticity is discussed in Sec.~\ref{sec:unit}, in
particular, we discuss the dispersive representation of the scalar pion
form factor and show how dispersion relations and CHPT can be 
combined to give accurate predictions of low-energy observables.
A few selected applications that represent the state-of-the-art in CHPT
and a discussion of the interplay between lattice QCD (LQCD) and CHPT
are given in Sec.~\ref{sec:appl}.

\section{QCD symmetries and their realization}
\label{sec:symm}

First, we must discuss chiral symmetry in the context of QCD. Chromodynamics is a
non-abelian $SU(3)_{\rm color}$ gauge theory with $N_f$ flavors of quarks,
three of them being light ($u,d,s$) and the other three heavy ($c,b,t$).
Here, light and heavy refers to a typical hadronic scale of about 1~GeV.
In what follows, we consider light quarks only (the heavy quarks are to be 
considered as decoupled). The QCD Lagrangian reads
\beq\label{LQCD}
{\mathcal L}_{\rm QCD} = -\frac{1}{2g^2} {\rm Tr} \left( G_{\mu\nu}G^{\mu\nu}
\right) + \bar q \, i \gamma^\mu D_\mu \, q -  \bar q {\mathcal M} \, q
={\mathcal L}_{\rm QCD}^0  -  \bar q {\mathcal M} \, q   ~,
\eeq
where we have absorbed the gauge coupling in the definition of the gluon
field and color indices are suppressed.  The three-component vector $q$
collects the quark fields, $q^T (x) = \left( u(s), d(x), s(x)\right)$. 
As far as the strong interactions
are concerned, the different quarks $u,d,s$ have identical properties, except
for their masses. The quark masses are free parameters in QCD - the theory
can be formulated for any value of the quark masses. In fact, light quark QCD
can be well approximated by a fictitious world of massless quarks,
denoted ${\mathcal L}_{\rm QCD}^0$ in Eq.~(\ref{LQCD}). Remarkably,
this theory contains no adjustable parameter - the gauge coupling $g$ merely
sets the scale for the renormalization group invariant scale $\Lambda_{\rm
  QCD}$. Furthermore, in the massless world left- and right-handed quarks
are completely decoupled. These are defined via
\beq
q_L = P_L \, q ~, \quad q_R = P_R \, q~, 
\eeq
in terms of the projection operators
\beqa
P_L &=& \frac{1}{2} (1 + \gamma_5)~,~~ P_R = \frac{1}{2} (1 - \gamma_5)~,
\nonumber\\
P_L^2 &=& P_L~,~~P_R^2 = P_R~,~~ P_L + P_R = {\bf 1}~, ~~ P_L \cdot P_R = 0~.
\eeqa
The Lagrangian of massless QCD is invariant under separate unitary
global transformations of the left- and right-hand quark fields, the
so-called {\em chiral rotations},
\beq
q_I \to V_I q_I~, \quad V_I \in U(3)~, \quad I = L,R~,
\eeq
leading to $3^2 =9$ conserved left- and $9$ conserved right-handed 
currents by virtue of Noether's theorem. These can be expressed in terms of 
vector ($V \sim L + R$) and  axial-vector ($A \sim L - R$) currents
\beqa\label{conscurr}
V_0^\mu &=& \bar q \, \gamma^\mu \, q~, \quad 
V_\mu^a = \bar q \, \gamma^\mu \frac{\lambda_a}{2} \, q~,\nonumber\\
A_0^\mu &=& \bar q \, \gamma^\mu \gamma_5 \, q~,\quad 
A_\mu^a = \bar q \, \gamma^\mu \gamma_5 \frac{\lambda_a}{2} \, q~,
\eeqa
Here, $a = 1, \ldots 8$, and the $\lambda_a$ are Gell-Mann's $SU(3)$
flavor matrices. As discussed later, the singlet axial current is anomalous,
and thus not conserved. The actual symmetry group of massless QCD
is generated by the charges of the conserved currents, it is
$G_0 = SU(3)_R \times SU(3)_L \times U(1)_V$. The $U(1)_V$ subgroup of 
$G_0$ generates conserved baryon number since the isosinglet vector 
current counts the number of quarks minus antiquarks in a hadron. 
The remaining group $SU(3)_R \times SU(3)_L$ is often
referred to as chiral $SU(3)$. Note that one also considers the light
$u$ and $d$ quarks only (with the strange quark mass fixed at its
physical value), in that case, one speaks of chiral $SU(2)$ and
must replace the generators in Eq.~(\ref{conscurr}) by the Pauli-matrices.
Let us mention that QCD is also invariant under the discrete symmetries of
parity ($P$), charge conjugation ($C$) and time reversal ($T$). 
Although interesting in itself, we do not consider strong $CP$ violation and
the related $\theta$-term in what follows, see e.g.~\cite{Peccei:1977hh}.

The chiral symmetry is a symmetry of the Lagrangian of QCD but not of the
ground state or the particle spectrum -- to describe the strong interactions
in nature, it is crucial that chiral symmetry is spontaneously broken. This
can be most easily seen from the fact that hadrons do not appear in parity
doublets. If chiral symmetry were exact, from any hadron one could generate
by virtue of an axial transformation another state of exactly the same 
quantum numbers except of opposite parity. The spontaneous symmetry breaking
leads to  the formation of a quark condensate in the vacuum $\sim 
\langle 0 | \bar q q|0\rangle =\langle 0  | \bar q_L q_R + \bar q_R q_L|0\rangle$, 
thus connecting the left- with the right-handed
quarks. In the absence of quark masses this expectation value
is flavor-independent: $\langle 0 | \bar u u|0\rangle = 
\langle 0 | \bar d d|0\rangle = \langle 0 | \bar q q|0\rangle$. 
More precisely, the vacuum is only invariant under the subgroup of 
vector rotations times the baryon number current, $H_0 = SU(3)_V \times
U(1)_V$. This is the generally accepted picture that is supported by general
arguments \cite{Vafa:1983tf} as well as lattice simulations of QCD (for 
a recent study, see~\cite{McNeile:2005pd} and references therein).
In fact, the vacuum expectation value of the quark condensate is only one
of the many possible order parameters characterizing the spontaneous symmetry
violation - all operators that share the invariance properties of the
vacuum (Lorentz invariance, parity, invariance under $SU(3)_V$
transformations) qualify as order parameters. The quark condensate
nevertheless enjoys a special role, it can be shown to be related to the
density of small eigenvalues of the QCD Dirac operator 
(see \cite{Banks:1979yr} and more recent discussions 
in \cite{Leutwyler:1992yt,Stern:1998dy}),
\beq
\lim_{{\cal M} \to 0} \langle 0 | \bar q q|0\rangle = - \pi \, \rho(0)~.
\eeq
For free fields, $\rho (\lambda) \sim \lambda^3$ near $\lambda = 0$. Only if
the eigenvalues accumulate near zero, one obtains a non-vanishing condensate.
This scenario is indeed supported by lattice simulations and many model studies
involving topological objects like instantons or monopoles.

Before discussing the implications of spontaneous symmetry breaking for QCD,
we briefly remind the reader of Goldstone's theorem 
\cite{Goldstone:1961eq,Goldstone:1962es}: to every generator of
a spontaneously broken symmetry corresponds a massless excitation of the
vacuum. This can be understood in a nut-shell (ignoring subtleties like
the normalization of states and alike - the argument also goes through in 
a more rigorous formulation). Be ${\cal H}$ some Hamiltonian that is invariant
under some charges $Q^i$, i.e. $[{\cal H}, Q^i] = 0$ with $i = 1, \ldots, n$.
Assume further that $m$ of these charges ($m \leq n$) do not annihilate the
vacuum, that is $Q^j |0 \rangle \neq 0$ for $j =1, \ldots, m$.  Define a
single-particle state via $|\psi\rangle = Q^j |0 \rangle$.   This is an
energy eigenstate with eigenvalue zero, since $H |\psi\rangle  = H
Q^j|0\rangle=  Q^j H
|0\rangle = 0$. Thus, $|\psi\rangle$ is a single-particle state with $E =
\vec{p} = 0$, i.e. a massless excitation of the vacuum. These states are the
{\em Goldstone bosons}, collectively denoted as pions $\pi(x)$ in what follows.
Through the corresponding symmetry current the Goldstone bosons couple 
directly to the vacuum,
\beq\label{GBME}
\langle 0 | J^0 (0) | \pi \rangle \neq 0~.
\eeq
In fact, the non-vanishing of this matrix element is a {\it necessary and
sufficient} condition for spontaneous symmetry breaking. In QCD, we have
eight (three) Goldstone bosons for $SU(3)$ ($SU(2)$) with spin zero and 
negative parity -- the latter property is a consequence that these Goldstone 
bosons are generated by applying the axial charges on the vacuum. The
dimensionful scale associated with the matrix element Eq.~(\ref{GBME}) 
is the pion decay constant (in the chiral limit)
\beq\label{GBM}
\langle 0|A^a_\mu(0)|\pi^b(p)\rangle = i \delta^{ab} F p_\mu~,
\eeq
which is a fundamental mass scale of low-energy QCD.
In the world of massless quarks, the value of $F$ differs from the
physical value by terms proportional to the quark masses, to be
introduced later, $F_\pi = F [1 + {\cal O}({\cal M})]$. The physical
value of $F_\pi$ is $92.4\,$MeV, determined from pion decay, $\pi\to \nu\mu$. 
For a discussion of $F_\pi$ in the context of the SM and beyond, see~\cite{Stern:2006ru}.

Of course, in QCD the quark masses are not exactly zero. The quark mass term leads
to the so-called {\em explicit chiral symmetry breaking}. Consequently, the
vector and axial-vector currents are no longer conserved (with the exception
of the baryon number current)
\beq\label{div}
\partial_\mu V_a^\mu = \dfrac{1}{2} i \bar q \, [{\cal M},\lambda_a ]\, q~, \quad
\partial_\mu A_a^\mu = \dfrac{1}{2} i \bar q \, \{{\cal M},\lambda_a \} \, \gamma_5\, q~.
\eeq
However, the consequences of the spontaneous symmetry violation  can still be
analyzed systematically because the quark masses are {\em small}. QCD
possesses what is called an approximate chiral symmetry. In that case, the mass spectrum
of the unperturbed Hamiltonian and the one including the quark masses can not be
significantly different. Stated differently, the effects of the explicit symmetry
breaking can be analysed in perturbation theory. This perturbation generates
the remarkable mass gap of the theory - the pions (and, to a lesser extent,
the kaons and the eta) are much lighter than all other hadrons.  To be more
specific, consider chiral $SU(2)$. The second formula of Eq.~(\ref{div}) is
nothing but a Ward-identity (WI) that relates the axial current $A^\mu =  \bar d
\gamma^\mu \gamma_5 u$ with the pseudoscalar density $P = \bar d i \gamma_5
u$,
\beq
\partial_\mu A^\mu = (m_u + m_d)\, P~.
\eeq
Taking on-shell pion matrix elements of this WI, one arrives at
\beq\label{pimass}
M_\pi^2 = (m_u + m_d) \frac{G_\pi}{F_\pi}~,
\eeq
where the coupling $G_\pi$ is given by $\langle 0 | P(0)| \pi(p)\rangle =
G_\pi $. This equation leads to some intriguing consequences: In
the chiral limit, the pion mass is exactly zero - in accordance with Goldstone's
theorem. More precisely, the ratio $G_\pi /F_\pi$ is a constant in the chiral
limit and the pion mass grows as $\sqrt(m_u+m_d)$ as the quark masses are
turned on. A more detailed discussion of the Goldstone boson masses and their
relation to the quark masses will be given in Section~\ref{sec:GBmass}.

There is even further symmetry related to the quark mass term. It is observed
that hadrons appear in isospin multiplets, characterized by very tiny
splittings of the order of a few MeV. These are generated by the small
quark mass difference $m_u -m_d$ (small with respect to the
typical hadronic mass scale of a few hundred MeV)
and also by electromagnetic effects of the
same size (with the notable exception of the charged to neutral pion mass
difference that is almost entirely of electromagnetic origin). This can be
made more precise: For $m_u = m_d$, QCD is  invariant under $SU(2)$ isospin 
transformations: 
\beq q \to q' = U q~,~~ q = \left(\begin{array}{cc} u \\ d\end{array}\right)~,
~~~U = \left(\begin{array}{cc} a^* &  b^* \\ -b & a\end{array}\right)~,
~~~|a|^2+|b|^2 = 1~.
\eeq
In this limit, up and down quarks can not be disentangled as far as the
strong interactions are concerned.  Rewriting of the QCD quark mass term
allows to make the strong isospin violation explicit:
\beq
{\cal H}_{\rm QCD}^{\rm SB} = m_u \,\bar u u + m_d
  \,\bar  d d = \dfrac{1}{2} (m_u+m_d)(\bar u u + \bar d d)
              + \dfrac{1}{2} (m_u-m_d)(\bar u u - \bar d d)~,
\eeq
where the first (second) term is an isoscalar (isovector). Extending 
these considerations to $SU(3)$, one arrives at the eighfold way of
Gell-Mann and Ne'eman \cite{GMNbook}
that played a decisive role in our understanding
of the quark structure of the hadrons. The $SU(3)$ flavor symmetry
is also an approximate one, but the breaking is much stronger than it is
the case for isospin. From this, one can directly infer that the quark mass
difference $m_s - m_d$ must be much bigger than $m_d -m_u$. Again, this
will be made more precise in Section~\ref{sec:GBmass}.

There is one further source of symmetry breaking, which is best understood
in terms of the path integral representation of QCD. The effective
action contains an integral over the quark fields that can be expressed
in terms of the so-called fermion determinant. Invariance of the action 
under chiral transformations not only requires the action to be left
invariant, but also the fermion measure \cite{Fujikawa:1983bg}. Symbolically,
\beq
\int [d\bar q][dq] \ldots \to |{\mathcal J}|\int [d\bar q'][dq'] \ldots
\eeq   
If the Jacobian is not equal to one, $|{\mathcal J}| \neq 1$,
one encounters an {\it anomaly}. Of course, such a statement has to be 
made more precise since the path  integral requires regularization and 
renormalization, still it captures the essence of the chiral anomalies
of QCD. One can show in general that certain 3-, 4-, and 5-point
functions with an odd number of external axial-vector sources are 
anomalous. As particular examples we mention the famous triangle
anomalies of Adler, Bell and Jackiw and the divergence of the
singlet axial current,
\beq
\partial_\mu (\bar q \gamma^\mu \gamma_5 q) = 2iq m \gamma_5 q +
\frac{N_f}{8\pi} G_{\mu\mu}^a \tilde{G}^{\mu\mu, a}~,
\eeq
that is related to the generation of the $\eta '$  mass. There are
many interesting aspects of anomalies in the context of QCD and chiral
perturbation theory. Space does not allow to discuss these, we refer
to~\cite{Bijnens:1993xi}. 

We end this section by giving list of reviews on the foundations and
applications of CHPT, see
\cite{Donoghue:1990pf,Bijnens:1993xi,Meissner:1993ah,Leutwyler:1994fi,Ecker:1994gg,Bernard:1995dp,Pich:1995bw,Gasser:1999fb,Leutwyler:2000jg,Scherer:2002tk,Gasser:2003cg}
and a  recent status report is Ref.~\cite{Ecker:2005za}. The
state-of-the-art two--loop calculations are reviewed in Ref.~\cite{Bijnens:2006zp}.

\section{Effective chiral Lagrangian and power counting}
\label{sec:Leff}

The appropriate work-horse to analyze the consequences of the
spontaneous, the explicit and the anomalous symmetry breaking in
QCD is the chiral effective Lagrangian \cite{Weinberg:1978kz}.
The relevant degrees of freedom are the Goldstone bosons coupled 
to external fields. Two remarks are in order: i) extensions of this 
scheme to include e.g. matter fields are briefly discussed below, 
and ii) one can equally well work with the generating functional, see
e.g. the classical papers \cite{Gasser:1983yg,Gasser:1984gg}. In the
chiral limit, we are dealing with a theory {\it without mass gap}.
Consequently, S-matrix elements and transition currents are dominated
by pion-exchange contributions. The QCD Lagrangian can thus be mapped onto
an effective Lagrangian,
\beq
{\mathcal L}_{\rm QCD} [\bar q , q, G] \to 
{\mathcal L}_{\rm eff} [U, \partial_\mu U, \ldots, {\mathcal M}]~,
\eeq
where $U(x)$ is a  matrix-valued $SU(3)$ field that collects the Goldstone
bosons. Further, $\partial_\mu U$ reminds us that all interactions are 
of derivative nature due to Goldstone's theorem and ${\mathcal M}$
keeps track of the explicit symmetry breaking due to the finite quark masses.
A formal proof of this equivalence based on the analysis of the chiral
Ward identities has been given by Leutwyler~\cite{Leutwyler:1993iq}
and by Weinberg~\cite{Weinberg:1994tu} --- space does not allow to discuss 
these beautiful papers in more detail here. The effective Lagrangian
leads to a well defined quantum field theory in which gauge and
chiral symmetries as well as the chiral anomaly are manifest. The best strategy
to construct the most general ${\cal L}_{\rm eff}$ consistent with the
QCD symmetries is to consider QCD in the presence of locally chiral invariant
external fields,
\beqa
{\mathcal L}_{\rm QCD} &=& {\mathcal L}_{\rm QCD}^0 
- \bar q \, \gamma_\mu  (v^\mu + \gamma_5 a^\mu) \, q
+ \bar q \,  (s - i \gamma_5 p) \, q~, \nonumber \\
&=& {\mathcal L}_{\rm QCD}^0 
- \bar{q}_L \, \gamma_\mu P_L l^\mu \, q_L
- \bar{q}_R \, \gamma_\mu P_R r^\mu \, q_R + \ldots
\eeqa
in terms of scalar $s(x)$, pseudoscalar $p(x)$, vector $v^\mu(x)$ and
axial--vector $a^\mu(x)$ sources. Explicit symmetry breaking is 
included in the scalar source, $s(x) = {\mathcal M} + \ldots =
{\rm diag}(m_u.m_d,m_s) + \ldots$. Electroweak interactions are
easily incorporated via
\beqa
r_\mu &=& e \, Q\, {\cal A}_\mu~, ~~ l_\mu = e \, Q\, {\cal A}_\mu
+ \frac{e}{\sqrt{2} \sin^2 \theta_W}\left( W_\mu^+ T_+ + {\rm h.c.}
\right)~, \nonumber\\
Q &=& {\rm diag}\left( \dfrac{2}{3},  -\dfrac{1}{3},  -\dfrac{1}{3} \right)~,
~~T_+ = \left(\begin{array}{ccc} 0 & V_{ud} & V_{us} \\
                                 0 &  0     &  0     \\
                                 0 &  0     &  0     \end{array}\right)~, 
\eeqa
with ${\cal A}_\mu$ the photon field, $W_\mu$ is the charged massive vector boson field,
$\theta_W$ the weak mixing angle and
$V_{ud}, V_{us}$ are the pertinent elements of the CKM matrix.  The most important
ingredient to make the effective field theory a useful tool is the {\bf power
counting}. In CHPT, we have a dual expansion in small external momenta and
small quark masses (with a fixed ratio to have a well defined chiral limit).
The corresponding small parameter is denoted by $q$, where small refers to the
typical hadronic scale of about 1~GeV. First, one assigns a chiral dimension
to all building blocks of ${\mathcal L}_{\rm eff}$: $U(x) = {\cal O}(1)$, 
$\partial_\mu U(x), l_\mu (x), r_\mu (x) = {\cal O}(q)$ and $s(x), p(x) =
{\cal O}(q^2)$. The last assignment is a consequence of Eq.~(\ref{pimass}) ---
the Goldstone boson masses are non--analytic in the quark masses. 
(For an alternative power counting, see~\cite{Fuchs:1991cq}).
The lowest order effective Lagrangian then takes the form
\beq\label{eq:L2}
{\mathcal L}^{(2)} = \frac{F^2}{4} \langle D_\mu U D^\mu U^\dagger 
+ \chi U^\dagger + \chi^\dagger U \rangle~,
\eeq
where the brackets denote the trace in flavor space,
$D_\mu U = \partial_\mu U + i l_\mu U - i U r_\mu$ is the chiral covariant
derivative and $\chi = 2 B(s+ip)$ parameterizes the explicit chiral 
symmetry breaking. The Lagrangian Eq.~(\ref{eq:L2}) is consistent with
the strictures from Goldstone's theorem.
To this order, the theory is completely specified by 
two parameters, the pion decay constant in the chiral limit $F$, 
cf. Eq.~(\ref{GBM}), and  $B$ measures the strength of the quark condensate 
in the chiral limit,  $B = |\langle 0|\bar q q|0\rangle|/F^2$.
At next-to-leading order ${\cal O}(q^4)$, the effective Lagrangian contains
10 (7)  local operators for $SU(3)\, (SU(2))$. These are accompanied by coupling
constants not determined by chiral symmetry, the so-called low-energy 
constants (LECs). For the explicit form of ${\mathcal L}^{(4)}$,
see~\cite{Gasser:1983yg,Gasser:1984gg}. However, at this order there
are further contributions. Interactions generate loops, e.g. closing two
external lines in a tree-level pion-pion scattering graph leads to the
one-loop pion tadpole (pion mass shift)
and the chiral anomaly is formally of order $q^4$. At two loop order,
one has further contributions from tree graphs with dimension six insertions,
from one-loop graphs with exactly one insertion from ${\cal L}^{(4)}$ and
two--loop graphs with insertions from ${\cal L}^{(2)}$. The complete structure
of the effective Lagrangian at two loop order is given in Ref.~\cite{Bijnens:1999sh}
(where one can also find references to earlier work on that topic).
All this is captured in the power counting formula of 
Weinberg~\cite{Weinberg:1978kz}, which orders the various contributions 
to any S--matrix element for pion interactions according to the
chiral dimension $D$ (the inclusion of external fields is straightforward),
\beq
D = 2 + \sum_d N_d (d-2) + 2 L~,
\eeq
with $N_d$ the number of vertices with dimension $d$ (derivatives and/or pion
mass insertions) and $L$ the number of pion loops. Chiral symmetry gives a
lower bound for $D$, $D \geq 2$ -- these are exactly the tree graphs with
lowest order $d =2$ vertices and $L = 0$ (giving the soft-pion (current
algebra) predictions of the sixties).

To address issues like isospin violation or the extraction of quark mass
ratios, one must include virtual photons and leptons in the EFT. Space forbids 
to discuss the many interesting aspects of these extensions, the interested
reader might consult some of the classics, see Refs.
\cite{Gasser:1982ap,Urech:1994hd,Meissner:1997fa,Knecht:1999ag,Schweizer:2002ft,Gasser:2003hk}.

Matter fields like e.g.\ nucleons can also be included in chiral perturbation
theory. In that case, special care has to be taken of the new (hard) mass
scale introduced by the matter field (such as the nucleon mass in the
chiral limit). This can be treated in various ways
for baryons (heavy fermion approach, infrared regularization, extended
on-mass-shell scheme and so on). A detailed review on this topic is 
Ref.~\cite{Bernard:1995dp} and more recent updates can be found in 
Refs.~\cite{Meissner:2000gh,Scherer:2005rm,Bernard:200X}. Virtual photons
in baryon CHPT are addressed e.g. in Refs.\cite{Muller:1999ww,Gasser:2002am}.

\section{Chiral loops and low-energy constants}
\label{sec:loops}

Beyond tree level, any observable calculated in CHPT receives contributions
from tree and and loop graphs. The loops not only generate the imaginary
parts  but are also -- in most cases -- divergent requiring regularization and 
renormalization. In CHPT,  one usually chooses a mass--independent
regularization scheme to avoid power divergences (there are, however,
instances where other regulators are more appropriate or physically intuitive.
For a beautiful discussion of this and related issues,
see e.g. Refs.~\cite{Georgi:1994qn,Espriu:1993if}). 
The method of choice in CHPT
is dimensional regularization (DR), which introduces the scale $\lambda$. 
Varying this scale has no influence on any
observable $O$ (renormalization scale invariance),
\beq
\dfrac{d}{d\lambda} \, O (\lambda) = 0~,
\eeq
but this also means that it makes little sense to assign a physical meaning to
the separate contributions from the contact terms and the loops. Physics,
however, dictates the range of scales appropriate for the process under
consideration --- describing the pion vector radius (at one loop) 
by chiral loops alone would 
necessitate a scale of about 1/2~TeV (as stressed long ago by Leutwyler). In
this case, the coupling of the $\rho$--meson generates the strength of the
corresponding one-loop counterterm that gives most of the pion radius ---
more on this below. The most intriguing aspects of chiral loops are the
so-called {\bf chiral logarithms} (chiral logs). In the chiral limit, the pion 
cloud becomes long-ranged and there is no more Yukawa factor $\sim \exp(-M_\pi
r)$ to cut it off. This generates terms  like $\log M^2_\pi, 1/ M_\pi,
\ldots$, that is contributions that are non--analytic in the quark masses.
Such statements can be applied to all hadrons that are surrounded by a cloud of pions 
which by virtue of their small masses can move away very far from the object
that generates them. Stated differently, in QCD the approach to the chiral limit
is non--analytic in the quark masses and the low--energy structure of QCD can
therefore not be analyzed in terms of a simple Taylor expansion. 
(An early paper that deals with the subtleties of approaching the chiral limit 
in QCD is Ref.~\cite{Gasser:1979hf}  and literature quoted therein).
The exchange of the massless Goldstone bosons generates poles and cuts 
starting at zero momentum transfer, such that the Taylor series expansion 
in powers of the momenta fails. This is a general phenomenon of theories 
that contain massless particles -- the Coulomb scattering amplitude due to 
photon exchange is proportional to $e^2/t$, with $t = (p'-p)^2$ the momentum 
transfer squared between the two charged particles.

As stated before, most loops are divergent. In DR, all one--loop divergences
are simple poles in $1/(d-4)$, where $d$ is the number of space-time
dimensions (for renormalization at two loops, see e.g. Ref.~\cite{Bijnens:1999hw}).
Consequently, these divergences can be absorbed in the pertinent LECs,
\beq
L_i \to L_i^{\rm ren} + \beta_i \, {\rm L}(\lambda)~, ~~
{\rm L}(\lambda) = \frac{\lambda^{d-4}}{16\pi^2} \left( \dfrac{1}{d-4} -
\frac{1}{2}\left( \ln(4\pi) + \Gamma^\prime (1) + 1 \right) \right)~,
\eeq
with $\beta_i$ the corresponding $\beta$--function and the renormalized
and finite  $L_i^{\rm ren}$ must be determined by a fit to data
(or calculated eventually using lattice QCD). Having determined the values of
the LECs from experiment, one is faced with the issue of trying to understand
these numbers? Not surprisingly, the higher mass states of QCD leave their
imprint in the LECs. Consider again the $\rho$-meson contribution to the
vector radius of the pion. Expanding the $\rho$-propagator in powers of
$t/M_\rho^2$, its first term is  a contact term of dimension four, with
the corresponding finite LEC $L_9$ given by $L_9 = F_\pi^2/2M_\rho^2 \simeq 7.2 \cdot
10^{-3}$, close to the empirical value $L_9 = 6.9  \cdot 10^{-3}$ at $\lambda
= M_\rho$. This so--called resonance saturation 
(pioneered in Refs.\cite{Ecker:1988te,Ecker:1989yg,Donoghue:1988ed})
 holds more generally for most LECs at one loop and is frequently used 
in two--loops calculations to estimate the ${\cal O}(p^6)$ LECs (for a
recent study on this issue, see~\cite{Kampf:2006bn}).
More
precisely, there are two types of LECs --- the so-called dynamical LECs
and the symmetry breakers. The contributions proportional to the  LECs 
of the first type are non--vanishing
in the chiral limit and can be determined from phenomenology. The symmetry
breakers, however, are much more difficult to pin down from data and are
also difficult to model. Here, recent progress in lattice QCD promises
a determination of the contribution from these LECs at unphysical 
values of the quark masses.
Much progress has been made in the field of resonance saturation
in the last years, for a 
state-of-the-art calculation see~\cite{Cirigliano:2006hb} (and the many
references therein). For
extensions of the idea of resonance saturation of the LECs in the pion--nucleon 
and the two--nucleon sectors, see e.g. Refs.~\cite{Bernard:1996gq,Epelbaum:2001fm}. 

Let us end this section with a short remark on the pion cloud of the
nucleon, a topic that has gained some prominence in recent years - the literature
abounds with incorrect statements. Consider as an example the isovector
Dirac radius of the proton \cite{Bernard:2003rp}.
At third order in the chiral expansion,
it takes the form
\beq\label{eq:Dirac}
\langle r^2 \rangle_1^V = \left(0.61 - \left(0.47\,{\rm GeV}^{-2}\right)
\tilde{d} (\lambda) + 0.47 \log\frac{\lambda}{1\,{\rm GeV}} \right) ~{\rm
fm}^2~, 
\eeq
where $\tilde{d} (\lambda)$ is a dimension three pion--nucleon LEC that
parameterizes the ``nucleon core'' contribution.  Comparing
Eq.~(\ref{eq:Dirac}) with the empirical value for the Dirac radius,
$\langle r^2 \rangle_1^V = (0.585\,{\rm fm})^2$,
one finds that even the sign of the core contribution 
$\sim \tilde{d} (\lambda)$ is not fixed if
$\lambda$ is varied within the sensible range from 600~MeV to 1~GeV.
Only the sum of the core and the cloud contribution constitutes a 
meaningful quantity that should be discussed.

\section{A specific one--loop calculation}
\label{sec:sff}

\begin{figure}[t!]
\centerline{\epsfysize=4cm\epsfbox{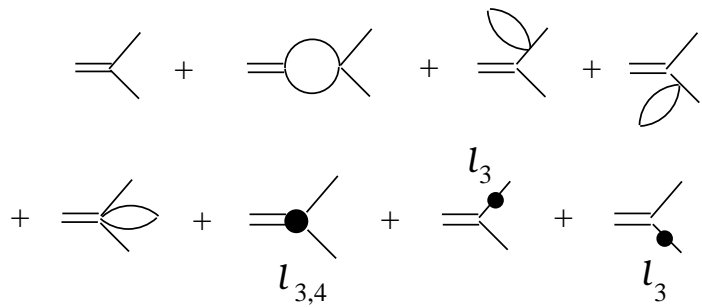}}
\caption{Graphs contributing to the pion scalar form factor at one loop.
The double line denotes the scalar--isoscalar source, solid lines are pions.
The filled circle depicts an insertion from the next-to-leading order
effective Lagrangian.}
\label{fig:sff}
\end{figure}

Let us now consider a specific example of a one--loop calculation based
on Feynman diagrams -- the scalar form factor of the pion 
(the same calculation using the generating functional
can be found in the classical paper~\cite{Gasser:1983yg}). 
We consider this simple 3-point function because it allows to make our
arguments with the least amount of algebra. In chiral $SU(2)$,
the coupling of the pion to a scalar-isoscalar source defines the scalar 
form factor $F_S (t)$,
\beq 
\delta^{ik} \, F_S(t) = \langle \pi^i(p') | \bar q q| \pi^k (p)\rangle~,
\quad t = (p'-p)^2~,
\eeq
with $i,k$ isospin indices and $t$ the invariant four-momentum transfer
squared. Since there are no scalar--isoscalar sources, this form factor
can only be indirectly inferred making use e.g. of dispersive techniques
(see Sec.~\ref{sec:unit} or Ref.~\cite{Donoghue:1990xh}). 
The important role of the scalar form factor 
stems from the observation  that its value at $t = 0$  is proportional to
the expectation value of the quark mass term in the QCD Hamiltonian,
\beq
\frac{\partial M_\pi^2}{\partial \hat{m}} = \langle \pi | \bar qq|\pi\rangle~.
\eeq 
To one--loop accuracy, one finds (the pertinent
tree and one--loop graphs are shown in Fig.~\ref{fig:sff}) 
\beqa\label{eq:sffone}
f(t) &=& 1 + h(t) + {\mathcal O}(p^4)~,\nonumber\\
h(t) &=& h_0 + h_1 \, t + \frac{1}{2F_\pi^2} \, \left( 2t - M_\pi^2 \right)
\, \bar{J} (t)~.
\eeqa
Here, $f(t) = F_S(t)/2B$ 
is the normalized scalar form factor and 
\beq
\bar{J} (t) = \frac{1}{16\pi^2} \left( \sigma 
\ln \frac{\sigma -1}{\sigma + 1}+2\right)~, \quad \sigma = 
\sqrt{1 - \frac{4M_\pi^2}{t}}~~~~~~ (t < 0)~,
\eeq
is the fundamental meson loop-integral (the so-called fundamental bubble) and
$h_0$ and $h_1$ are polynoms in the pion mass (modulo logs) 
that depend on  the one--loop 
renormalized LECs $\ell_3$ and $\ell_4$,
\beqa\label{eq:sffpoly}
h_0 &=& \frac{M_\pi^2}{16 \pi^2 F_\pi^2}\, \left( \ln\frac{M_\pi^2}{\mu^2} +
64 \pi^2 \,\ell_3(\mu) + \frac{1}{2} \right)~,\nonumber\\
h_1 &=& \frac{1}{16 \pi^2 F_\pi^2} \, \left(- \ln\frac{M_\pi^2}{\mu^2} +
16 \pi^2 \,\ell_4(\mu) - 1 \right)~.
\eeqa
These LECs are universal and relate various Green functions. For the case
at hand, $\ell_4$ can be obtained from the ratio $F_K/F_\pi$ (extending the
theory to $SU(3)$ and then matching the corresponding LEC to $\ell_4$ by
integrating out the kaons and the eta) and $\ell_3$
from the expansion of $M_\pi$ in powers of the quark masses. The chiral 
logarithms in $h_0$ and $h_1$ are generated by some of the loop graphs
depicted in Fig.~\ref{fig:sff}. Note that the scalar
form factor is finite in the chiral limit. It is instructive to study its 
expansion at low momentum transfer,
\beq
f(t) = 1 + \frac{1}{6} \langle r_S^2 \rangle \,t + {\mathcal O}(t^2)~,
\eeq
which defines the scalar radius $r_S$. Its low-energy representation can be
read off from Eqs.~(\ref{eq:sffone},\ref{eq:sffpoly})
\beq
\langle r_S^2 \rangle = 6h_1 - \frac{1}{192\pi^2F_\pi^2}~.
\eeq
Remarkably, the scalar radius is
sizably larger than the corresponding vector radius (that can  be
extracted from $e^+ e^- \to \pi^+ \pi^-$ data, see e.g.~\cite{Bijnens:2002hp})
\beq
\langle r_S^2 \rangle = (0.61 \pm 0.02)~{\rm fm}^2 \gg 
\langle r_V^2 \rangle = (0.452 \pm 0.013)~{\rm fm}^2~,
\eeq
with the values for the scalar radius taken from~\cite{Donoghue:1990xh}.
This difference is understood - it is generated by the strong pion--pion
interaction in the isospin zero S-wave. This can also be seen from the fact
that the coefficient of the chiral loga\-rithm contained in $\langle r_S^2
\rangle$ is six times larger than the time-honored corresponding coefficient in
$\langle r_V^2 \rangle$~\cite{Beg:1973sc}.

\section{Exploring analyticity and unitarity}
\label{sec:unit}

In CHPT, imaginary parts of vertex functions and scattering amplitudes are
generated by loop diagrams so that unitarity is obeyed perturbatively.
The non--trivial unitarity effects generated by the loops are generated
by the propagation of  on-shell intermediate states in the loop diagrams. 
Causality implies certain properties of the analytic structure of amplitudes
that allow one to relate real and imaginary parts in form of dispersion
relations. Consider e.g. the dispersion relation for the normalized scalar form factor,
\begin{equation}
f(s) = \frac{1}{\pi} \, \int_0^\infty ds' \frac{{\rm Im}~f(s')}{s-s'-i\epsilon}~.
\end{equation}
Knowledge of  ${\rm Im}~f(s)$ for all $s$ thus allows to reconstruct $f(s)$
in the low energy region (and above). It is evident that subtractions of the
dispersion relation can soften the dependence of this integral on large $s$.
The contents of the chiral loops and the content of the dispersive integral
must therefore be related, even more, possible subtraction constants in the
dispersive representation must be mapped onto combinations of LECs since they
represent the most general polynomial contribution consistent with the
underlying symmetries. It was therefore argued early (see e.g.
Refs.~\cite{Truong:1987ex,Donoghue:1993kw,Donoghue:1995pd})
 that dispersion 
relations might be used to extend the range of applicability of CHPT. However,
one has to make the relation between the chiral and the dispersive
representations more precise.
 \begin{figure}[t!]
\centerline{\epsfysize=4cm\epsfbox{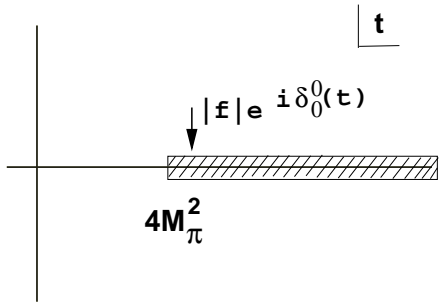}}
\caption{Analytic properties of the scalar form factor.}
\label{fig:cutff}
\end{figure}
\noindent
To appreciate the content of the dispersive compared to the chiral 
representation, consider again the normalized scalar form factor of the 
pion, $f(t)$. It is given in terms of an 
analytic function in the complex $t$-plane, cut along the real axis for 
$t \leq  4M_\pi^2$, cf. Fig.~\ref{fig:cutff}:
\beq
\frac{1}{2i} \left[ f(t + i \epsilon) - f(t-i\epsilon)\right] 
= |f(t)| \, \sin \delta_0^0(t)~,
\eeq
making use of Watson's theorem, $f(t) = |f(t)| \exp(i\delta_0^0(t))$, with 
$\delta_0^0$ the elastic $\pi\pi$ isospin zero, S-wave scattering 
phase shift. This holds to very good approximation up to the $\bar KK$ threshold. 
At order $p^2$, the scalar form factor has a discontinuity given by
\beq
\delta_0^0(t) = \frac{1}{32\pi^2 F_\pi^2} \sqrt{1-\frac{4M_\pi^2}{t}} 
\, \left(2t - M_\pi^2\right)~.
\eeq
We now want to construct a function that has exactly this discontinuity,
\beqa
k(t) &=& \frac{1}{2F_\pi^2} \, (2t - M_\pi^2)\, \bar{J} (t)  ~,
\eeqa
so that
\beq\label{eq:ffdisp}
f(t) = a + b \,t  +  \frac{1}{2F_\pi^2} \, (2t - M_\pi^2)\, \bar{J} (t)  ~,
\eeq
with $a$ and $b$ unknown coefficients. However, we are able to draw some 
conclusions on the pion mass dependence of these coefficients. Because the 
form factor is normalized to one at $t = 0$, we know that
\beq\label{eq:acond}
a = 1 + a_1 M_\pi^2 ~~~ ({\rm modulo}~~{\rm logs})~.
\eeq
As the pion mass tends to zero, we can expand the function $\bar{J}(t)$,
\beq
\bar{J}(t) \stackrel{M_\pi \to 0}{\longrightarrow} \frac{1}{16\pi^2} \ln
M_\pi^2 + \ldots~,
\eeq
where the ellipsis denotes terms not relevant for the following. In order
to have a finite scalar form factor in the chiral limit, we must require
\beq\label{eq:bcond}
b = - \frac{1}{16\pi^2} \ln M_\pi^2~.
\eeq
Combining Eqs.~(\ref{eq:ffdisp},\ref{eq:acond},\ref{eq:bcond}) we have obtained a
representation that is algebraically
equivalent to the one--loop representation given
in Eq.~(\ref{eq:sffone}) --- without having calculated a single loop diagram. 
This is truly a pleasure. The procedure we have performed to relate the
LECs with the subtraction constants is referred to as {\bf matching}. Matching
can be done at various orders. At tree level, the form factor is real and one 
thus only needs to ensure that the normalization is correct, see
Eq.~(\ref{eq:acond}). Matching at the one--loop level allows one to reconstruct
the chiral representation at that order, with the subtraction constants taking
over the role of certain LECs. It is instructive to take a closer look at this
dispersive representation of $f(s)$ as compared to the chiral one. First,
note that the two results have the same algebraic structure, the dispersive 
representation contains, however, less information. The subtraction 
constants $a$ and $b$ take the role
of the LECs $\ell_3$ and $\ell_4$ in Eq.(\ref{eq:sffpoly}) 
-- but such subtraction constants are process-dependent and can not be 
related to other Green functions. Nevertheless, the chiral log in $b$ 
lets one understand the enhancement of the corresponding LEC. 
These arguments can easily be carried out to higher orders -- the dispersive 
representation of the scalar form factor based on the one--loop CHPT amplitude 
is worked out  in~\cite{Gasser:1990bv} and the full two--loop result was later given 
in~\cite{Bijnens:1998fm} (for more recent work on the scalar form factors of
the pion and the kaon, see e.g. 
Refs.~\cite{Moussallam:1999aq,Meissner:2000bc,Frink:2002ht,Ananthanarayan:2004xy,Lahde:2006wr}).
To summarize this little exercise, as long as one 
is interested in the algebraic structure of a given observable (matrix
element), the dispersive method is fine and also easy to apply. However, to
relate Green functions to other quantities or to analyze the {\it complete}
pion  mass dependence of observables or LECs, a one (or higher) loop
calculation in CHPT is mandatory.

Historically, the first use of analyticity and unitarity dates back
long before the advent of CHPT - namely the calculation of Lehmann
of massless pion-pion scattering to fourth order in the pion 
momenta~\cite{Lehmann:1972kv}. Since his arguments are so elegant,
it is worth repeating them here. For massless pions, the leading
order $\pi\pi$ scattering amplitude is given in terms of a single
invariant function, $A^{(2)}(s,t,u) = s/F^2$, with $F$ the pion decay constant
in the chiral limit and the Mandelstam variables obey $s+t+u = 0$. Further,
$A(s,t,u)$ must be symmetric in $t,u$ (Bose symmetry). Thus, at fourth order
one has only two independent combinations, $s^2 = (t+u)^2$ and
$tu$. This fixes the polynomial parts at fourth order. However, at this
order the scattering amplitude is no longer real. If one  employs elastic 
unitarity of the scattering amplitude, ${\rm Im}\,T = |T|^2$, it follows that 
the imaginary part of the invariant function $A^{(4)}$ takes the form
\beq 
{\rm Im}~A^{(4)} =  \frac{1}{16\pi F^4} \left\{ \frac{1}{2} s^2 \theta(s)
+ \frac{1}{6} t(t-u) \theta(t) + \frac{1}{6} u(u-t) \theta(u)\right\}~,
\eeq
where the three terms in the curly brackets are due to the $s$-, $t$- and
$u$-channel cuts, respectively. Next, one makes use of analyticity to
construct a function that produces this imaginary part - such a function is
\beq
A^{(4)}  = \frac{1}{16\pi^2 F^4} \left\{ - \frac{1}{2} s^2 \log\frac{-s}{s_0}
- \frac{1}{6} t(t-u) \log\frac{-t}{t_0} - \frac{1}{6} u(u-t) 
\log\frac{-u}{t_0} \right\}~,
\eeq
where $s_0$ and $t_0$ are constants and
in the last term $t_0$ appears because of Bose symmetry. This result is
quite remarkable - at fourth order the $\pi\pi$ scattering amplitude depends
solely on two constants that can not be determined from analyticity. If we now
introduce two scale-dependent parameters $G_1 (\mu^2)$ and $G_2 (\mu^2)$, we
obtain for the amplitude at fourth order;
\beqa
A(s,t,u) &=& A^{(2)} (s,t,u) + A^{(4)} (s,t,u) + {\mathcal O}(s^3, \ldots )
 \nonumber\\
A^{(4)} (s,t,u) &=& G_1 (\mu^2) \, s^2 + \, G_2 (\mu^2)\, tu \nonumber\\
&+& \frac{1}{16\pi^2 F^4} \Biggl\{ - \frac{1}{2} s^2 \log\frac{-s}{\mu^2} 
  - \frac{1}{6} t(t-u) \log\frac{-t}{\mu^2}
- \frac{1}{6} u(u-t) \log\frac{-u}{\mu^2} \Biggr\}~, \nonumber\\
G_1 (\mu) &=& \frac{1}{32\pi^2 F^4} \left( \log\frac{s_0}{\mu^2} + \frac{1}{3}
 \log\frac{t_0}{\mu^2} \right)~,\nonumber\\ 
G_2 (\mu^2)  &=& -\frac{1}{24\pi^2 F^4} \left( \log\frac{t_0}{\mu^2} \right)~. 
\eeqa
This dispersive representation can be matched to the one--loop CHPT
representation for massless pions which contains the two LECs $\ell_1$
and $\ell_2$ -- completely analogous to the case of the scalar form factor,
cf. Eq.~(\ref{eq:sffpoly}). 

In the meantime, this program has been carried
much further by combining the 
Roy equations~\cite{Roy:1971tc} -- special dispersion relations
utilizing the high degree of crossing symmetry of the elastic pion 
scattering amplitude  --
with the two--loop chiral representation for the $\pi\pi$ scattering
amplitude. This has lead to the remarkable prediction of 
Ref.~\cite{Colangelo:2000jc} for the S-wave scattering lengths,
\beq\label{eq:a0a2}
a_0^0 = 0.220 \pm 0.05~, \quad a_0^2 = -0.0444 \pm 0.0010~.
\eeq
Such a precision is rarely achieved in low-energy QCD.  Space forbids to
describe these wonderful calculations in detail, the interested reader
might consult the original literature --  
see Refs.~\cite{Knecht:1995tr,Bijnens:1995yn,Bijnens:1997vq}
for the two--loop representation of the $\pi\pi$ amplitude, 
see Ref.~\cite{Ananthanarayan:2000ht} for a review on
Roy equation studies of $\pi\pi$ scattering and 
Refs.~\cite{Descotes-Genon:2001tn,Kaminski:2006yv}
for other calculations of this type (see also \cite{Pelaez:2003eh}). 
The comparison of the
chiral prediction with experimental determinations of the scattering
lengths will be discussed in Sect.~\ref{sec:pipi}.

Another interesting consequence of unitarity are the so-called (threshold)
cusps that appear in scattering processes when a new channel opens. More
precisely, due to kinematical reasons such cusps are only visible in S-waves.
One example is the already discussed scalar form factor of the pion which
exhibits a cusp at the two-pion threshold, see e.g. Fig.~2
in~\cite{Gasser:1990bv}. A similar effect in the vector form factor is
washed out by the kinematical P-wave prefactors. Another cusp effect that
was long considered an academic curiosity appears in  $\pi^0\pi^0 \to \pi^0\pi^0$
scattering due to the opening of the $\pi^+ \pi^-$ threshold just 9.2~MeV
about the $\pi^0 \pi^0$ threshold, as first found in~\cite{Meissner:1997fa}
(and graphically shown in Ref.~\cite{Meissner:1997gf}). It was only realized  
years later that the same cusp effect appears in the decay $K^+ \to 
\pi^+ \pi^0 \pi^0$ and allows one to accurately extract the scattering
length combination $a_0^0 - a_0^2$ \cite{Cabibbo:2004gq}.
Further theoretical work on this issue can be found in 
Refs.\cite{Cabibbo:2005ez,Colangelo:2006va}. In the last reference, a consistent
field-theoretical method is developed to analyze this effect consistently.
For the present state of experimental determinations of  $a_0^0 - a_0^2$
from kaon decays see Sect.~\ref{sec:pipi}. Another manifestation of this
phenomenon appears in neutral pion production off the nucleon as discussed in
Sect.~\ref{sec:photo}.

\section{Applications}
\label{sec:appl}

\subsection{Goldstone boson masses} 
\label{sec:GBmass}

\begin{figure}[t!]
\vspace{8mm}
\centerline{\epsfxsize=10cm\epsfbox{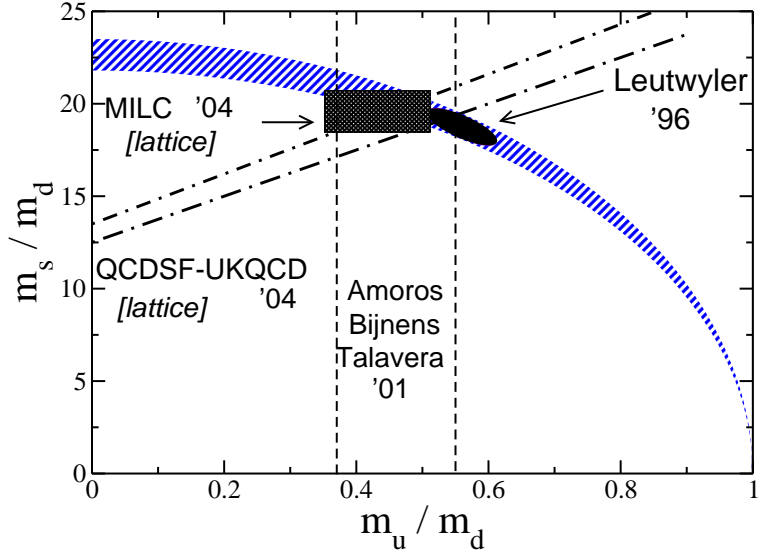}}
\caption{Light quark mass ratios. The light shaded band 
is the first quadrant of Leutwyler's ellipse for $Q = 22.7 \pm 0.8$.
Depicted are results from Leutwyler  \cite{Leutwyler:1996qg} 
and from the Lund group \cite{Amoros:2001cp} as well as from two recent
lattice calculations \cite{Aubin:2004fs,Gockeler:2004rp}.
Figure courtesy of J\"urg Gasser.
}
\label{fig:ellipse}
\end{figure}

One of the most interesting applications of the CHPT machinery is the
extraction  of the light quark mass {\em ratios} e.g. from the chiral
expansion of the Goldstone boson masses. Only with additional input,
say from QCD sum rules or lattice QCD, one is able to extract values
of the quark masses in a given scheme at a given scale -- in CHPT
the quark masses always appear together with the LEC $B$. The basic
ideas how to apply CHPT in the analysis of the quark mass ratios
and the state of the art of such extractions has been reviewed in this journal 
by Donoghue in 1989 \cite{Donoghue:1989sj}. Since then, theoretical activity
has focused on  Leutwyler's ellipse that relates the quark mass ratios
$m_s/m_d$ and $m_u/m_d$ via
\beq
\dfrac{1}{Q^2} \left(\frac{m_s}{m_d}\right)^2 +
 \left(\frac{m_u}{m_d}\right)^2 = 1
\eeq
modulo corrections of ${\cal O}(m_d^2/m_s^2)$, with
$Q^2 = (m_s^2-\hat{m}^2)/(m_d^2-m_u^2)$ and $\hat m = (m_u+m_d)/2$. 
The numerical value of $Q \simeq 23$ is difficult to pin down 
accurately because it sensitively depends on the
next-to-leading order electromagnetic corrections to the Goldstone
boson masses (the corrections to Dashen's theorem), see e.g. 
the discussion in \cite{Leutwyler:1996qg} and
references therein. Furthermore, in the effective field theory, 
a redefinition of the quark condensate and certain LECs allows 
one to freely move on the ellipse -- the famous Kaplan-Manohar 
ambiguity \cite{Kaplan:1986ru}. Historically, the ellipse was first
defined in the seminal work~\cite{Gasser:1984gg} and it was first drawn
in~\cite{Kaplan:1986ru}. It can also been shown that this ambiguity
persists at two--loop level, consequently some additional information
like e.g. quark mass ratios from the baryon mass splittings is needed
to pin down the physical values of the quark mass ratios. The state
of the art of such determinations is shown in Fig.~\ref{fig:ellipse}.
All determinations agree on one result - the up quark mass is non-zero
($m_u = 0$ would trivially solve the strong CP-problem).
\noindent
For orientation, we give here the results from~\cite{Leutwyler:1996qg}:
\beq
\dfrac{m_u}{m_d} = 0.553\pm 0.043~, ~~
\dfrac{m_s}{m_d} = 18.9\pm 0.8~, ~~
\dfrac{m_s}{\hat m} = 24.4\pm 1.5~.
\eeq
It is remarkable that the up and down quark masses are so different -- naively
one thus would expect very sizeable strong isospin violation. However, this 
difference is effectively masked because it is so small compared to any
hadronic scale, may it be $\Lambda_{\rm QCD}$, $m_\rho$ or $\Lambda_\chi$.
The role of $\bar ss$ fluctuations on the ratio $m_s/\hat m$ is discussed 
in~\cite{Descotes-Genon:2003cg}.

Another interesting recent result concerns the chiral expansion of the
pion mass, which to leading order is given by the Gell-Mann--Oakes--Renner
relation~\cite{Gell-Mann:1968rz}
\beq
M_\pi^2 = B \, (m_u + m_d) + {\cal O}(m_{u,d}^2)
\eeq
and the corrections quadratic in the quark masses are parameterized by one
LEC (called $\bar{\ell}_3$) that can also be obtained from data on $K_{\ell 4}$
decays (see next chapter).  
It was shown in Ref.~\cite{Colangelo:2001sp} that  $|\bar{\ell}_3|
\leq 16$, which implies that the  Gell-Mann--Oakes--Renner relation represents
a good approximation -- more than 94~\% of the pion mass originate from the
first term in its quark mass expansion. This also shows that the quark
condensate is indeed the leading order parameter.

\subsection{Goldstone boson scattering} 
\label{sec:pipi}

\begin{figure}[t!]
\vspace{9mm}
\centerline{\epsfxsize=10cm\epsfbox{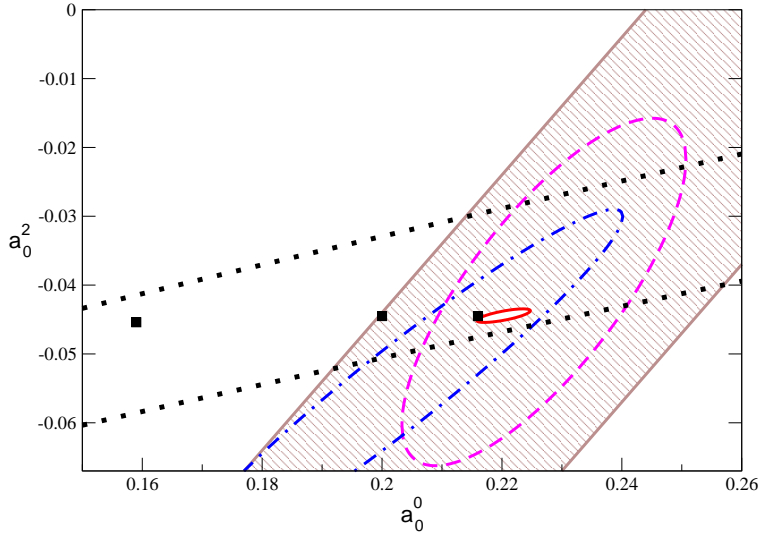}}
\caption{S-wave $\pi\pi$ scattering lengths: theory versus experiment.
The filled squares give the central value of the tree \cite{Weinberg:1966kf}, 
one-loop \cite{Gasser:1983kx} and two-loop~\cite{Bijnens:1995yn}
calculations, respectively. The small solid ellipse is the result of
Ref.~\cite{Colangelo:2000jc}. The dotted lines denote the universal
curve~\cite{Ananthanarayan:2000ht}.
The hatched band is the  results obtained from the
DIRAC experiment (pionium lifetime) \cite{Adeva:2005pg}, the dot-dashed ellipse
is the $K_{\ell4}$ result from E865 \cite{Pislak:2003sv}  and the dashed ellipse the
one obtained by the NA48/2 collaboration from analyzing the cusp 
in $K \to 3\pi$ \cite{Batley:2005ax}. Figure courtesy of Heiri Leutwyler.
}
\label{fig:a0a2}
\end{figure}

The purest reaction to test the chiral dynamics of QCD is elastic
pion--pion scattering at threshold. Since the pion three--momentum
vanishes at threshold, the dual expansion of CHPT is given by one
single small parameter, $M_\pi^2 / (4\pi F_\pi)^2 \simeq 0.014$.
The chiral expansion for the S-wave scattering lengths takes the
form
\beqa
a_0^0 = \frac{7 M_\pi^2}{32 \pi F_\pi^2}~ \Bigl[ 1 + \Delta_4 + \Delta_6
\Bigr] + {\mathcal O} (M_\pi^8)~, \nonumber\\
 a_0^2 = -\frac{M_\pi^2}{16 \pi F_\pi^2}~ \Bigl[ 1 + 
\tilde\Delta_4 + \tilde\Delta_6 \Bigr] + {\mathcal O} (M_\pi^8)~,
\eeqa
where $\Delta_4$ and  $\Delta_6$ collect the one-- and two--loop corrections,
first given in \cite{Gasser:1983kx} and \cite{Bijnens:1995yn}, respectively.
The numerical evaluation of these corrections gives 
as central values (using $F_\pi = 92.4\,$MeV) 
\beq
a_0^0 = 0.159 [ 1 + 0.26 + 0.10] = 0.216~, ~~
a_0^2 = 0.0454 [ 1 - 0.02 + 0.00] = 0.0445~.
\eeq
The corrections in the isospin zero channel are remarkably large - this
effect is completely  understood in terms of the very strong final-state
interactions that effectively generate a very broad pole at $\sqrt{s} \simeq
440\,$MeV far off the real axis. 
A layman's discussion of this effect is
provided in~\cite{Meissner:1990kz}. Quite in contrast, for isospin two
the corrections have the expected small size. As remarked earlier, the
most precise determination of these fundamental quantities of QCD
comes from a combination of CHPT with dispersion relations,
cf. Eq.~(\ref{eq:a0a2}). Fig.~\ref{fig:a0a2} collects the presently 
available theoretical and experimental information on the S-wave scattering
lengths. The universal curve was already established from dispersion theoretical
studies of $\pi\pi$ scattering in the sixties - it was found that all
solutions that give phase shifts compatible with the $\rho$--meson and 
Regge behavior at high energies lead to a narrow band for $a_0^0$ and $a_0^2$
\cite{Morgan:1969ca}. Shown is the updated universal curve from
\cite{Ananthanarayan:2000ht}.
The experimental information stems from the analysis of 
$K_{\ell 4}$ decays \cite{Pislak:2003sv}, the measurement of the lifetime of 
pionium (an electromagnetic bound state of a $\pi^+ \pi^-$ pair)  \cite{Adeva:2005pg}
and the analysis of the aforementioned cusp in $K\to 3\pi$ decays  \cite{Batley:2005ax}.
Note that the preliminary analysis of $K_{e4}$ data from NA48/2 seems to be in
conflict with these values.

The next simple and pure Goldstone boson process including also strange quarks is
elastic pion--kaon scattering in the threshold region. Here, the situation is
less satisfactory. First, the expansion parameter is now  
$M_K^2 / (4\pi F_\pi)^2 \simeq 0.18$, which is sizably larger than in the
$SU(2)$ case. The one-- and two--loop corrections have been worked out in
\cite{Bernard:1990kx,Bernard:1990kw} and \cite{Bijnens:2004bu}, respectively.
Furthermore, the Roy-Steiner equations for pion-kaon scattering with
constraints from CHPT have  been developed and analyzed 
in Refs.~\cite{Ananthanarayan:2001uy,Buettiker:2003pp}.  It was found that
most of the low--energy data are only in poor agreement with the solutions of
the Roy--Steiner equations. Nevertheless, for the S-wave  scattering lengths
in the basis of physical isospin 1/2 and 3/2, the chiral expansion seems to 
converge reasonably well. A typical two--loop result from 
Ref.~\cite{Bijnens:2004bu} is  (note, however, that the paper also 
contains other fits with very different values)
\beq 
a_0^{1/2} = 0.220 ~[0.224 \pm 0.022]~, ~~~
10 \, a_0^{3/2} = -0.47 ~[-0.448 \pm 0.077]~,
\eeq
where the numbers in the brackets are the dispersive results of 
Ref.~\cite{Buettiker:2003pp}. However, there persists a real puzzle.
One can formulate an $SU(2)$ low-energy theorem for the isovector
scattering length $a_0^- = (a_0^{1/2} - a_0^{3/2})/3$ \cite{Roessl:1999iu},
\beq\label{eq:piK}
a_0^- = \dfrac{M_\pi^2}{8\pi F_\pi^2 (1 +M_\pi/M_K)} \left( 1 + {\cal O}
(M_\pi^2) \right)
\eeq
which is not affected by kaon loop effects at next-to-leading order.
(Note that there is a small caveat in this statement since the pertinent
LECs have not been properly adapted from $SU(3)$ to $SU(2)$.)
Since the final-state interactions in $K \pi$ scattering are weaker than
in $\pi \pi$, one expects smaller corrections to $a_0^-$ than to $a_0^0$.
This is also borne out by the one--loop calculation, the corrections is
about 12~\% \cite{Roessl:1999iu,Kubis:2001bx}. However, matching the
$SU(3)$ two--loop representation of  \cite{Bijnens:2004bu} to $SU(2)$, it was found
that the subleading corrections to the low--energy theorem Eq.~(\ref{eq:piK})
are of the same size as the leading ones
\cite{Schweizer:2005nn}. This might be related to the fact that a poor
convergence was also found for some of the subthreshold parameters that
parameterize the $\pi K$ scattering amplitude inside the Mandelstam triangle,
see  \cite{Bijnens:2004bu}. More work is needed to clarify the situation.

\subsection{Neutral pion photoproduction} 
\label{sec:photo}

\begin{figure}[t!]
\vspace{8mm}
\centerline{\psfig{file=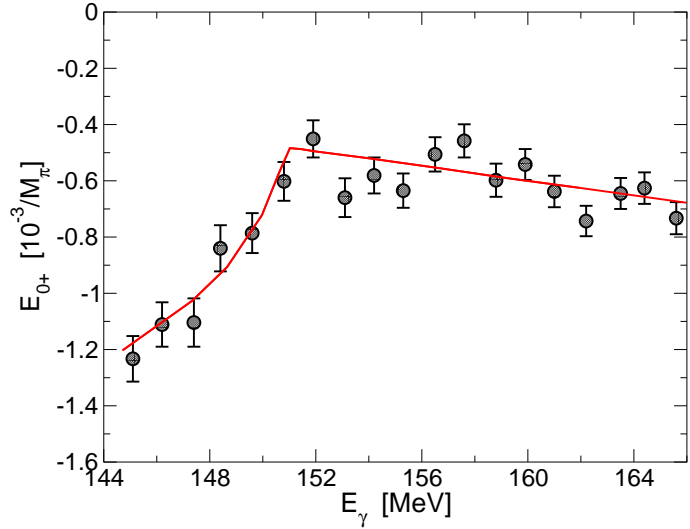,angle=270,width=9.0cm}}
\caption{The real part of the electric dipole amplitude $E_{0+}$ in the threshold
  region. The solid line is the one--loop CHPT prediction
  \cite{Bernard:1994gm,Bernard:2001gz} and the data are
  the most recent measurement from MAMI~\cite{Schmidt:2001vg}.  
}
\label{fig:e0p}
\end{figure}

The chiral structure of QCD can also be analyzed in the presence of matter
fields, in particular nucleons. Neutral pion photoproduction off nucleons in the 
threshold region, $\gamma N \to \pi^0 N$ $(N = p,n)$, exhibits one of the
most intriguing realizations of pion loop effects. In the threshold region,
this process can be parameterized in terms of one complex S--wave and three
complex P--wave multipoles, called $E_{0+}$ and $P_{1,2,3}$, respectively (for
precise definitions, see e.g. \cite{Bernard:1994gm} and references therein). 
To disentangle these multipoles, one has to measure differential cross sections and one
polarization observable, like e.g. the photon asymmetry in $\vec \gamma p \to
\pi^0 p$. Historically, a low-energy theorem (LET) for the electric dipole 
amplitude was derived in the heydays of current algebra under certain analyticity
(smoothness) assumptions~\cite{DeBaenst:1971hp,Vainshtein:1972ih}. Measurements at Mainz,
Saclay and Saskatoon seemed to be in conflict with this LET.  It was, however, 
shown in Ref.~\cite{Bernard:1991rt} that the presence of the pions in loop
graphs (more precisely in the so-called triangle diagram)  generates infrared 
singularities in the Taylor coefficients of the invariant amplitude for
$E_{0+}$ -- invalidating the smoothness assumption made in 
Refs.~\cite{DeBaenst:1971hp,Vainshtein:1972ih}. The correct form of the LET
thus reads
\beqa
\left\{ E_{0+}^{\pi^0 p} \atop  E_{0+}^{\pi^0 n} \right\} &=& -\dfrac{e \,
  g_{\pi N}}{8 \pi m} \left[ \mu\, F_1 + \mu^2\, F_2 + {\cal O}(\mu^3) \right]~,
\nonumber\\
F_1 = \left( 1 \atop 0 \right)~, ~~F_2 &=& \!\!\!
-\dfrac{1}{2} \left( 3+\kappa^p 
\atop -\kappa^n \right) - \Delta \, \left(1 \atop 1\right)~,~~
\Delta = \dfrac{m^2}{16 F_\pi^2} \simeq 6.4~,
\eeqa
in terms of the small parameter $\mu = M_\pi/ m \simeq 1/7$ and $m$ is the nucleon
mass. Furthermore, $g_{\pi N} \simeq 13$ is the strong pion--nucleon coupling
constant and $\kappa_{p,n}$ the anomalous magnetic moment of the proton and
neutron, respectively. The contribution $\sim \Delta$ is the novel pion
loop effect first found in~\cite{Bernard:1991rt}. In fact, the terms $\sim
\mu^3$ have also been worked out, besides loop contributions one has two
polynomial terms which in the threshold region can be combined in one 
structure accompanied by one LEC. Consider now for definiteness neutral pion
production off the proton. Fixing this LEC at the opening of the
$\pi^+ n$ threshold just $\approx 7\,$MeV above the $\pi^0 p$ threshold,
one obtains an excellent description of the threshold data for Re~$E_{0+}$
as shown in Fig.~\ref{fig:e0p}. Also clearly visible is the cusp at the
$\pi^+ n$ threshold --- its strength is  directly proportional to the isovector
(charge exchange) scattering length. Thus a more precise measurement of this
cusp would give additional experimental information on zero energy
pion--nucleon scattering (see also Ref.~\cite{Bernstein:1996vf} for further discussion).
\noindent
Also, CHPT predicts the counterintuitive result that at threshold
$|E_{0+}^{\pi^0 n}| >  |E_{0+}^{\pi^0 p}|$. This prediction is vindicated
by the measurement of coherent neutral pion production off the deuteron
and utilizing CHPT for few--nucleon systems, see~\cite{Beane:1997iv}.
Note, however, that the chiral expansion for the electric dipole amplitude
is not converging very well -- this is another manifestation of strong
final-state interactions, this time in the $\pi N$ system. Quite unexpectedly,
in Ref.~\cite{Bernard:1994gm} novel LETs for the P--wave multipoles $P_{1,2}$
were found (including terms of order $\mu$) and the corrections of ${\cal
  O}(\mu^2)$ were analyzed in~\cite{Bernard:2001gz}. These LETs are in
good agreement with the data from Mainz (which performed so far the
only polarization measurement that allows to disentangle $P_1$ from $P_2$), 
see e.g. \cite{Schmidt:2001vg} or \cite{Bernard:2001gz} for detailed
discussions. The comparison of the CHPT predictions with the most recent
data from Mainz reads:
\beqa\label{thrval}
E_{0+} &=& -1.19 \,~~[-1.23 \pm 0.08 \pm 0.03]~,\nonumber \\
\bar P_1 &=& \phantom{-}9.67 \,~~ [9.46  \pm 0.05 \pm 0.28]~,\nonumber\\
\bar P_2 &=& -9.6 \,~~~ [-9.5  \pm 0.09 \pm 0.28]~,
\eeqa 
in the conventional units of $10^{-3}/M_{\pi^+}$ and $10^{-3}/M_{\pi^+}^2$, 
respectively.
Note also that the third P--wave multipole $P_3$ is given in terms of one
third order LEC. $P_3$ is completely dominated by the excitation of the 
$\Delta$ resonance - in fact free fits to the data and estimating the strength
of the LEC from resonance saturation give almost identical results.
Recently, the relation between dispersion relations and the chiral
representation for neutral pion photoproduction has been analyzed in the
context of the so-called Fubini-Furlan-Rosetti sum rule, 
see Refs.~\cite{Pasquini:2004nq,Bernard:2005dj,Pasquini:2006yi}.
In particular, relations between the LECs and the subtractions constants of
the dispersion relations were derived and used to pin down some of the LECs
with an unprecedented accuracy - cf. the discussion in Sec.~\ref{sec:unit}.
For similar studies combining dispersion relations and CHPT for elastic
pion-nucleon scattering, see e.g. Refs.~\cite{Buettiker:1999ap,Becher:2001hv}.
In particular, in Ref.~\cite{Becher:2001hv} the Roy-type equations for $\pi N$
scattering are written down --- it just remains to solve them.

\subsection{Connection to lattice QCD} 
\label{sec:latt}

QCD matrix elements can  be calculated from first principles
in the framework of lattice QCD. Space-time is approximated by a box
of the size $V = L_s \times L_s \times L_s\times L_t$, where $L_s\,(L_t)$ 
refers to the length in the spatial (temporal) directions. Furthermore, an UV cutoff is
given by the inverse lattice spacing $a$. At present, typical lattice sizes 
and spacings are $L_s \simeq L_t \simeq 2\ldots 3\,$fm and $a \simeq 0.07 \ldots
0.15\,$fm, respectively.  In addition, it is very difficult
to perform numerical simulations with light fermions, so the state-of-the-art
calculations using various (improved) actions and modern algorithms
have barely reached pion masses of about 250~MeV. In addition, exact chiral
symmetry can only be implemented in the computer-time intensive  
domain wall \cite{Kaplan:1992bt} or overlap  \cite{Narayanan:1994gw} formulations.
Most results obtained so far are based on simulations using considerably 
heavier pions (and no exact chiral symmetry). To connect lattice
results to the real world of continuum QCD, one has to perform the continuum
limit ($a \to 0$), the thermodynamic limit $(V \to \infty)$ and chiral
extrapolations from the unphysical to the physical quark masses.
All this can be performed in the framework of suitably tailored effective
field theories, which are variations of continuum chiral perturbation theory 
discussed so far, like e.g. staggered CHPT, Wilson CHPT and so on, via the
intermediate step of the Symanzik effective action \cite{Symanzik:1983dc,Symanzik:1983gh}. 
Space does not allow to review all these interesting 
developments, we refer the reader to the recent comprehensive review by
Sharpe~\cite{Sharpe:2006pu} (for an early review on the CHPT treatment
of finite volume effects, see~\cite{Meissner:1993ah}). 
Obviously, the lattice practioneers need CHPT -- for truly ab initio
calculations the quark mass dependence of the lattice results must be
analyzed in terms of a {\em model-independent} approach like CHPT, in
a regime of quark masses where it is applicable. Using resummation schemes
or models to try to extend the range of applicability of CHPT inevitably
induces an uncontrolled systematic uncertainty that should be avoided.
It is indeed astonishing how often one finds in the literature statements
of precise determinations of hadron properties based on extrapolation
functions that are not rooted in QCD or are at best models with  a
questionable relation to QCD.
In what follows, we will address the issue to what extent chiral perturbation 
theory representations and  lattice QCD results are already overlapping. 

\begin{figure}[t!]
\vspace{8mm}
\centerline{\psfig{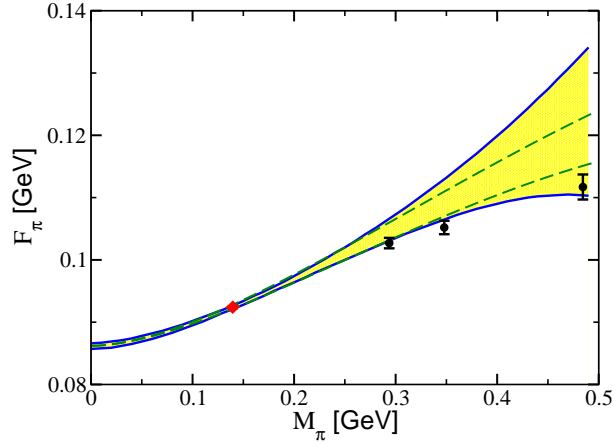}}
\caption{Pion (quark) mass dependence of the pion decay constant -- lattice
results from Ref.~\cite{Beane:2005rj} (circles) in comparison to the two--loop 
CHPT result from Ref.~\cite{Colangelo:2003hf}
(hatched area). The one--loop  band which is obtained by varying
$\tilde\ell_4$ is given by the dashed lines.
The diamond is the physical point.
}
\label{fig:Fpi}
\vspace{-5mm}
\end{figure}

Chiral perturbation theory provides unambiguous chiral extrapolation
functions, parameterized in terms of the pertinent LECs. Ideally, one
would like to perform global fits to a large variety of observables
since these are interrelated through the appearance of certain LECs.
At present, however, this is not yet possible and for a variety
of applications it is mandatory to include phenomenological input 
for some of the LECs -- an example will be given below. It should be
noted that frequently in the literature one--loop extrapolation functions
are used for pion masses well outside the regime of their applicability. 
As a matter of fact, most pion and kaon Green functions have been 
worked out at two--loop accuracy in the continuum (for a review, 
see~\cite{Bijnens:2006zp}) and considerable progress has been reported for 
many of these quantities for partially quenched QCD, in which one allows for 
different values of the valence and the sea quarks, see e.g.
\cite{Bijnens:2004hk,Bijnens:2005ae,Bijnens:2005pa,Bijnens:2006jv}.
Clearly, with increasing quark masses the CHPT representations become
increasingly inaccurate -- and this is in fact a strength of the effective
field theory, in that it provides a measure of the theoretical uncertainty.
Let us illustrate these issues for the pion decay constant $F_\pi$. Its special
role for the spontaneous symmetry breaking was already explained in 
Sec.~\ref{sec:symm}. Its analytic form at two-loop order in the continuum
is~\cite{Bijnens:1998fm,Colangelo:2003hf}
\beqa
F_\pi &=& F\, \left[ 1 + X \, \tilde\Delta^{(4)} + X^2 \, \tilde\Delta^{(6)} \right]
\nonumber\\
& =& F \, \biggl\{ 1 + X \,\left[ \tilde{L} +
\tilde{\ell}_4 \right] + X^2 \,\left[
-\frac{3}{4}\tilde{L}^2 + \tilde{L} \left( -\frac{7}{6}\tilde{\ell}_1
-\frac{4}{3}\tilde{\ell}_2 + \tilde{\ell}_4 - \frac{29}{12}\right)\right.
\nonumber\\
&& \qquad \qquad \qquad \qquad \qquad \qquad
\left. + \frac{1}{2}\tilde{\ell}_3\tilde{\ell}_4 + \frac{1}{12}
\tilde{\ell}_1 - \frac{1}{3}\tilde{\ell}_2 - \frac{13}{192} + \tilde{r} (\mu)
\right] \biggr\} ~, \nonumber\\
X & = & \frac{M_\pi^2}{16\pi^2 F^2} ~, ~~\tilde{L} =
\log\frac{\mu^2}{M_\pi^2}~,~~ \tilde{\ell}_i = \log\frac{\Lambda_i^2}{\mu^2}~,
\eeqa
where $\tilde{r} (\mu)$ is a combination of dimension six LECs. Note that --
in  contrast to common lore -- the number of new LECs does not explode when
going to higher orders when one considers specific observables.  With 
$\Lambda_1 = 0.12_{-0.03}^{+0.04}\,$GeV,
$\Lambda_2 = 1.20_{-0.06}^{+0.06}\,$GeV,
$\Lambda_3 = 0.59_{-0.41}^{+1.40}\,$GeV,
$\Lambda_4 = 1.25_{-0.03}^{+0.04}\,$GeV,
$\tilde{r} (\mu) = 0 \pm 3$ from resonance saturation, varying the scale of
dimensional regularization $\mu$ between 500~MeV and 1~GeV and using $F_\pi = 92.4
\pm 0.3\,$MeV, one obtains the (yellow) band between the solid lines 
in Fig.~\ref{fig:Fpi}. 
For comparison, the dashed lines correspond to 
the one--loop result. Note that
for pion masses below 300~MeV, the one-- and two--loop representations are
essentially equivalent, but for higher pion masses it is important to include
the two--loop corrections for a realistic assessment of the theoretical
uncertainty. This is consistent with expectations based on naive dimensional
analysis, the expansion parameter is $X = 0.01, 0.07, 0.18$ for $M_\pi =
139.57, 300, 500\,$MeV, respectively.
For orientation, we also show the recent lattice results from
 Ref.~\cite{Beane:2005rj}, in that paper, the one--loop representation for
$M_\pi$ and $F_\pi$ is used and the LEC $\tilde{\ell}_4$ is extracted with an
accuracy of a few percent. This appears optimistic if one accounts for the
two--loop corrections at these pion masses as shown in Fig.~\ref{fig:Fpi}. 
Other collaboration like QCDSF, ETM, CERN-Roma, $\ldots$ have also obtained 
results for $F_\pi$ at low pion masses, however, these ``data'' have not been 
finally analyzed by the time this review was written (see e.g. 
Ref.~\cite{ECTproc} for some of these  preliminary results).

Matters are somewhat different in the baryon (nucleon) sector for two reasons.
Although a multitude of ground-state (and some excited state) properties have
been simulated, most CHPT calculations have been performed to one--loop
accuracy. Further, in these extrapolation functions one has even and odd
powers of the expansion parameter(s) and correspondingly more LECs. For these
reasons chiral extrapolations can only be performed over a smaller range
of pion masses with a tolerable uncertainty as compared to the meson sector. 
Space does not allow for an overall review here, we refer the reader to 
Ref.~\cite{Bernard:200X}. We briefly discuss the quark mass dependence of the 
nucleon axial-vector coupling $g_A$ here, since a variety of lattice data 
exists for pion masses ranging from 340~MeV to 1~GeV. Also, the two--loop 
representation of $g_A$ has recently been published \cite{Bernard:2006te}. 
This allows one to discuss some subtleties that can arise in the chiral expansion 
of certain observables. 
The pion mass expansion of $g_A$ takes the form
\beq
g_A = g_0 \,\biggl\{ \underbrace{1}_{\rm tree} + 
\underbrace{\Delta^{(2)} + \Delta^{(3)}}_{\rm 1-loop} 
+ \underbrace{ \Delta^{(4)} + \Delta^{(5)}}_{\rm 2-loop} 
\biggr\}  + {\cal O}(M_\pi^6)~,
\eeq
where $g_0$ is the chiral limit value of $g_A$ and $\Delta^{(n)}$
collects the corrections that are proportional to $M_\pi^n$. At one--loop
order, the chiral representation of $g_A$ contains $g_0$,  one combination 
of dimension two LECs ($c_2+c_3$) and one
dimension three LEC ($d_{16}$). The latter two can be determined from the analysis
of $\pi N \to \pi N$ and $\pi N \to \pi \pi N$, respectively. This is one
of the cases where it is mandatory to include such phenomenological
information to analyze the chiral expansion of a given observable. 
The one--loop representation  is dominated by the $M_\pi^3$-term as 
the pion mass increases -- one is
therefore not able to connect to the lattice results, which show a very
weak dependence of $g_A$ on the pion mass. 
\begin{figure}[t!]
\vspace{8mm}
\centerline{\psfig{file=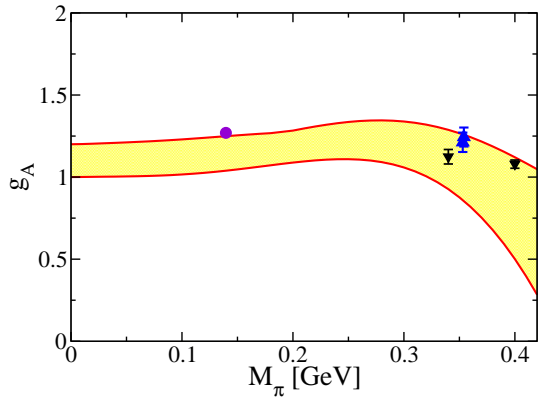,angle=270,width=8.0cm}}
\caption{Pion  mass dependence of the axial-vector coupling -- lattice
results from LHPC/MILC~\cite{Edwards:2005ym} (triangles) 
and QCDSF~\cite{QCDSFgA} (inverted triangles) 
in comparison to the two--loop CHPT result from Ref.~\cite{Bernard:2006te}
(filled area).  The circle is the physical point.
}
\vspace{-5mm}
\label{fig:gA}
\end{figure}
At two loops, one has further operators accompanied by certain
combinations of LECs. The dominant contributions to these stem from $1/m$
corrections to the lower order LECs. The remaining pieces can only be
estimated assuming naturalness. Demanding further that the trend of the
lattice data is followed, one arrives at the (yellow) band shown in
Fig.~\ref{fig:gA}. The width of the band is a consequence of this condition 
-- there are strong cancellations between the various contributions. Also, one sees
that the range of applicability of the chiral extrapolation function and the
lattice results barely overlap, to arrive at precision results for $g_A$,
pion masses of less than 300~MeV are mandatory. In view of this, the claim of 
Ref.~\cite{Edwards:2005ym} that $g_A$ has been calculated and precisely
determined in the chiral regime appears overly optimistic - more data at lower
pion masses are called for to substantiate such claims. We also note that
these strong cancellations between various orders were first found in an EFT
approach including the $\Delta (1232)$ as an active degree of freedom and
counting the nucleon-delta mass splitting $m_\Delta - m_N \simeq 3F_\pi$
as an additional small parameter at leading one--loop order,
see~\cite{Hemmert:2003cb} and the recent update~\cite{Procura:2006gq}.
The claims in these papers that one can truthfully represent the chiral
expansion of $g_A$ in leading one-loop order  for pion masses up to
700~MeV once the delta is included are unfounded -- the resummation of some
subset of corrections induces an uncontrolled uncertainty because other
higher order effects are ignored. A beautiful example for this is provided
by the cancellations of delta contributions and higher order $\pi N$ loop
effects in the nucleons' magnetic polarizabilities~\cite{Bernard:1993bg}.
Note also that finite volume corrections for $g_A$, which are essential for
connecting the lattice results with the real world, are discussed e.g. 
in Refs.~\cite{Beane:2004rf,Colangelo:2005cg,Khan:2006de}.

Finally, we note that CHPT practioneers also need the lattice. In particular
for non-leptonic weak meson interactions and in the baryon sector, the amount of
precise phenomenological information is limited. Therefore, it appears
impossible to pin down all LECs, in
particular the symmetry breakers, cf. Sec.~\ref{sec:loops}. After the
pioneering work of Ref.~\cite{Myint:1993hs} in the mid-nineties, the ALPHA
collaboration has reported considerable progress in the lattice determination
of the next-to-leading order LECs in the meson sector \cite{Heitger:2000ay}. 
Further work on the leading order LECs for $\Delta S =1$ transitions are reported in 
Ref.~\cite{Giusti:2004yp} and a method to determine the pion--nucleon LEC
$c_3$ was suggested in Ref.~\cite{Bedaque:2004dt}. In our opinion, more
effort should be invested in the determination of LECs from lattice studies,
and this would further tighten the interplay between the CHPT and the lattice
communities. This is important because it will take a long time before
complicated processes or reactions with many external probes will be amenable
to lattice simulations.


\section*{Acknowledgements} 
We are very grateful to J\"urg Gasser and Heiri Leutwyler for many useful
comments and communications. We have also profited from discussions with
Hans Bijnens, Gilberto Colangelo, John Donoghue, Gerhard Ecker, Bastian Kubis, 
Akaki Rusetsky and Jan Stern. 
Partial financial support under the EU Integrated Infrastructure
Initiative Hadron Physics Project (contract number RII3-CT-2004-506078),
by the DFG (TR 16, ``Subnuclear Structure of Matter'') and BMBF
(research grant 06BN411) is gratefully acknowledged.
This work was supported in part by the EU Contract No. MRTN-CT-2006-035482,
\lq\lq FLAVIAnet''.

\end{document}